\documentclass[final,3p,times,authoryear]{elsarticle}

\usepackage{euscript,amsmath,amssymb,amsfonts,graphicx,cite}
\allowdisplaybreaks[1]

\usepackage{chngcntr}
\counterwithout{figure}{section}
\counterwithout{equation}{section}
\usepackage{xcolor}

\usepackage{epstopdf}

\newcommand{\e}{{\rm e}}
\renewcommand{\d}{{\rm d}}

\newcommand{\ep}{\epsilon}

\newcommand{\D}{\displaystyle}
\newcommand{\mc}{\mathcal }

\renewcommand{\P}{{\rm P}}
\newcommand{\sign}{{\rm sign}}
\newcommand{\RR}{{\rm RR}}

\begin{document}
\begin{frontmatter}

\title{Optimizing sequential decisions in the drift-diffusion model}
\author[UH1]{Khanh P. Nguyen}
\address[UH1]{Department of Mathematics, University of Houston, Houston TX 77204 ({\tt kpnguyen@math.uh.edu}, {\tt josic@math.uh.edu})}
\author[UH1,UH2,Rice,eq]{Kre\v{s}imir Josi\'{c}}
\address[UH2]{Department of Biology and Biochemistry, University of Houston, Houston TX 77204}
\address[Rice]{Department of BioSciences, Rice University, Houston TX 77005}
\author[CU,SOM,eq]{Zachary P. Kilpatrick}
\address[CU]{Department of Applied Mathematics, University of Colorado, Boulder, Colorado 80309, USA ({\tt zpkilpat@colorado.edu}).}
\address[SOM]{Department of Physiology and Biophysics, University of Colorado School of Medicine, Aurora, CO 80045}
\address[eq]{equal authorship}

\begin{abstract}  To make decisions organisms often accumulate information across multiple timescales. However, most experimental and modeling studies of decision-making focus on sequences of independent trials. On the other hand, natural environments are characterized by long temporal correlations, and evidence used to make a present choice is often relevant to future decisions. To understand decision-making under these conditions we  analyze how a model ideal observer accumulates evidence to freely make choices across a sequence of correlated trials. We use principles of probabilistic inference to show that an ideal observer incorporates information obtained  on one trial as an initial bias on the next. This bias decreases the time, but not the accuracy of the next decision. Furthermore, in finite sequences of trials the rate of reward is maximized when the observer deliberates longer for early decisions, but responds more quickly towards the end of the sequence. Our model also explains experimentally observed patterns in decision times and choices, thus providing a mathematically principled foundation for evidence-accumulation models of sequential decisions.
\end{abstract}

\begin{keyword} 
decision-making, drift-diffusion model, reward rate, sequential correlations
\end{keyword}
\end{frontmatter}

%
%


\section{Introduction}

\label{intro}

Organismal behavior is often  driven by decisions that are the result of evidence  accumulated   to determine the best among available options~\citep{gold07,brody16}. For instance, honeybee swarms use a democratic process in which each bee's opinion is communicated to the group to decide which nectar source to forage~\citep{seeley91}. Competitive animals evaluate their opponents' attributes to decide whether to fight or flee~\citep{stevenson12}, and humans decide which stocks to buy or sell, based on individual research and social information~\citep{moat13}.  Importantly, the observations of these agents are frequently uncertain~\citep{hsu05,brunton13} so accurate decision-making requires robust evidence integration that accounts for the reliability and variety of evidence sources~\citep{raposo12}.

 The two alternative forced choice (TAFC) task paradigm has been successful in probing the behavioral trends and neural mechanisms underlying decision-making~\citep{ratcliff78}. In a TAFC task subjects decide which one of two hypotheses  is more likely based on  noisy evidence~\citep{gold02,bogacz06}. For instance, in the random dot motion discrimination task, subjects decide whether a cloud of noisy dots predominantly moves in one of two directions. Such stimuli evoke strong responses in primate motion-detecting areas, motivating their use in the experimental study of neural mechanisms underlying decision-making~\citep{shadlen01,gold07}. The response trends and underlying neural activity are well described by the drift-diffusion model (DDM), which associates a subject's belief with a particle drifting and diffusing between two boundaries, with decisions determined by the first boundary the particle encounters~\citep{stone60,bogacz06,ratcliff08}. 

The DDM is popular because (a) it can be derived as the continuum limit of the statistically-optimal sequential probability ratio test~\citep{wald48,bogacz06}; (b) it is an analytically tractable Wiener diffusion process whose summary statistics can be computed explicitly~\citep{ratcliff04,bogacz06}; and (c) it can be fit remarkably well to behavioral responses and neural activity in TAFC tasks with independent trials~\citep{gold02,gold07} (although see~\citet{latimer15}). 

However, the classical DDM does not describe many aspects of decision--making  in natural environments.  For instance, the DDM is typically used to model a series of independent trials where evidence accumulated during one trial is not informative about the correct choice on other trials~\citep{ratcliff08}.  Organisms in nature often  make a sequence of related decisions based on overlapping evidence~\citep{chittka09}. Consider an animal deciding which way to turn while fleeing a pursuing predator: To maximize its chances of escape its decisions should depend on both its own and the predator's earlier movements~\citep{corcoran16}. Animals foraging over multiple days are biased towards food sites with consistently high yields~\citep{gordon91}. Thus even in a variable environment, organisms use previously gathered information to make future decisions~\citep{dehaene12}. We need to extend previous experimental designs and corresponding models to understand if and how they do so.

Even in a sequence of independent trials, previous choices influence subsequent decisions~\citep{Fernberger20}. Such serial response dependencies have been observed in TAFC tasks which do~\citep{cho02,Frund14} and do not~\citep{bertelson61} require accumulation of evidence across trials. For instance, a subject's response time may decrease when the current state is the same as the previous state~\citep{pashler91}. Thus, trends in response time and accuracy suggest subjects use trial history to predict the current state of the environment, albeit suboptimally~\citep{kirby76}.  

\citet{goldfarb12}  examined responses in a series of dependent trials with the correct choice across trials evolving according to a two-state Markov process. The  transition probabilities affected the response time and accuracy of subjects in ways well described by a DDM with biased initial conditions and thresholds. For instance, when repeated states were more likely, response times decreased in the second of two repeated trials. History-dependent biases also increase the probability of repeat responses when subjects view sequences with repetition probabilities above chance~\citep{abrahamyan16,Braun18}. These results suggest that an adaptive DDM with an initial condition biased towards the previous decision is a good model of human decision-making across correlated trials~\citep{goldfarb12}.

Most earlier models proposed  to explain these observations are intuitive and recapitulate behavioral data, but are at least partly heuristic~\citep{goldfarb12,Frund14}. \citet{Yu08} have proposed a normative model that assumes the subjects are learning non-stationary transition rates in a dependent sequence of trails. However, they did not examine the  normative model for  known transition rates.  Such a model provides a standard for experimental subject performance, and a basis for approximate models, allowing us to better understand the heuristics subjects use to make decisions~\citep{ratcliff08,brunton13}. 

Here, we  extend previous  DDMs to provide a normative model of evidence accumulation in serial trials evolving according to a two-state Markov process whose transition rate is known to the observer. We use sequential analysis to derive the posterior for the environmental states $H_{\pm}$ given a stream of noisy observations~\citep{bogacz06,velizcuba16}. Under this model, ideal observers incorporate information from previous trials to bias their initial belief on subsequent trials.  This decreases the average  time to the next decision,  but not necessarily its accuracy. Furthermore, in finite sequences of trials the rate of reward is maximized when the observer deliberates longer for early decisions, but responds more quickly towards the end of the sequence. We also show that the model agrees with experimentally observed trends in decisions. 

\begin{figure}[t]
\begin{center} \includegraphics[width=10cm]{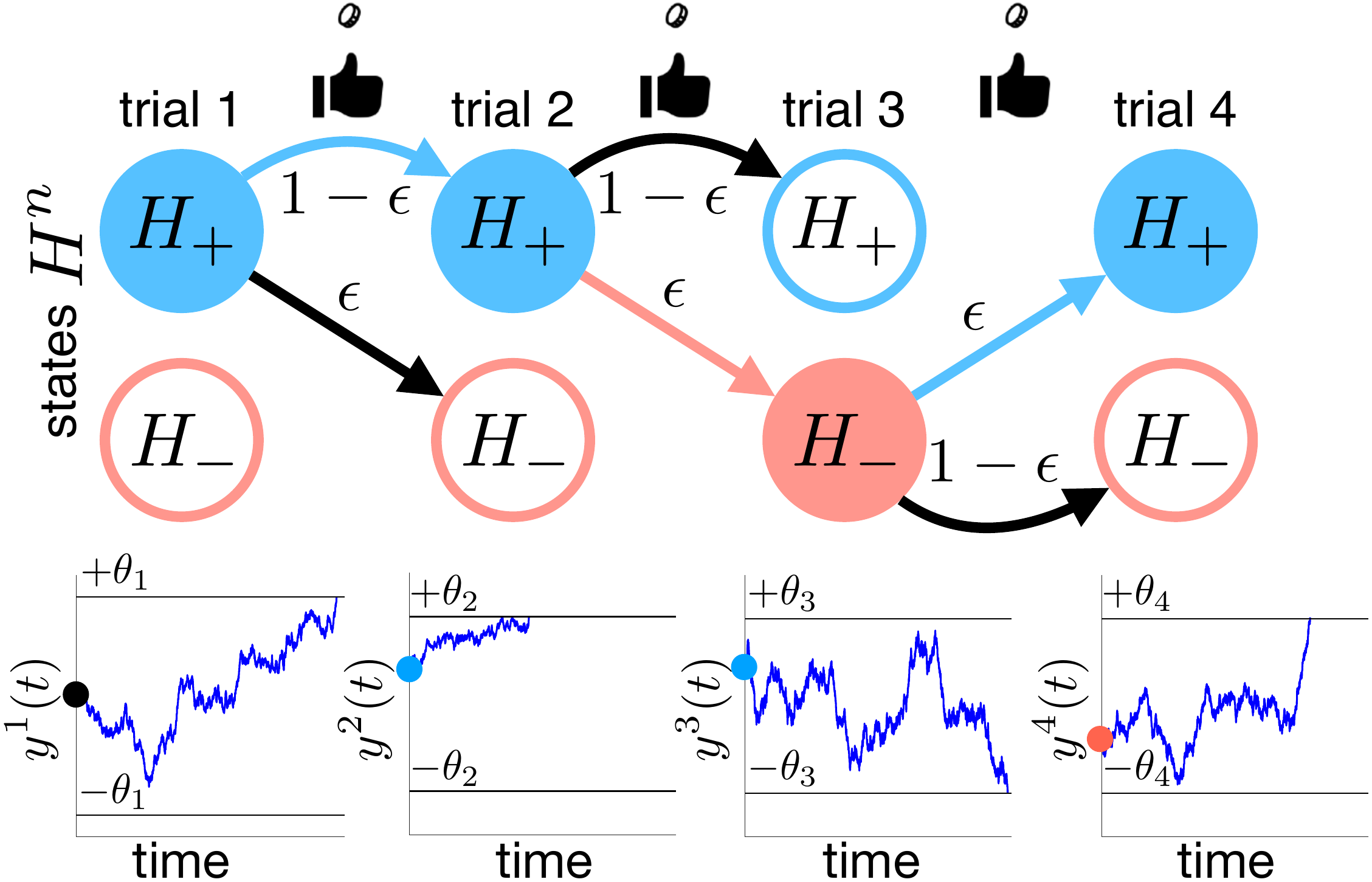} \end{center}
\vspace{0mm}
\caption{Sequence of TAFC trials with states $H^n$ determined by a two-state Markov process with switching probability $\ep := \P(H^n \neq H^{n-1})$. For instance, $H^{1:4} = (H_+,H_+,H_-, H_+)$ could stand for the drift direction of each drift-diffusion process $y^n(t)$ corresponding to the decision variable of an ideal observer making noisy measurements, $\xi^n(t)$. The decision threshold ($\pm \theta_{n-1}$) crossed in trial $n-1$ determines the sign of the initial condition in trial $n$: $y^n(0)$.}
\label{fig1:schematic}
\vspace{-4mm}
\end{figure}

\section{A model of sequential decisions}
\label{model}
We model a repeated TAFC task with the environmental state in each trial ($H_+$ or $H_-$) chosen according to a two-state Markov process: In a sequence of $n$ trials, the correct choices (environmental states or hypotheses) $H^{1:n} = (H^1, H^2, ...,H^n)$ are generated so that $P(H^1 = H_{\pm}) = 1/2$ and $P(H^{i} = H_{\mp} | H^{i-1} = H_{\pm}) = \ep$ for $i=2,...,n$ (See Fig.~\ref{fig1:schematic}).  
When $\ep = 0.5$ the states of subsequent trials are independent, and ideally the decision on one trial should not bias future decisions~\citep{ratcliff78,shadlen01,gold07,gold02,bogacz06,ratcliff08}.

When $0 \leq \ep < 0.5$, repetitions are more likely and trials are dependent, and evidence obtained during one trial can inform decisions on the next\footnote{If $0.5 < \ep \leq 1$ states are more likely to alternate. The analysis is similar to the one we present here, so we do not discuss this case separately.}. We will show that ideal observers use their decision on the previous trial to adjust their prior over the states at the beginning of the following trial.

Importantly, we assume that all rewards are given at the end of the trial sequence~\citep{Braun18}.
Rewarding the correct choice on  trial $i$ provides unambiguous information about the state $H^i$, superseding 
the noisy information gathered over that trial.  
Our results can be easily extended to this case, as well as cases when rewards are only given with some probability.

Our main contribution  is to derive and analyze the ideal observer model for this sequential version of the TAFC task. We use principles of probabilistic inference  to derive the sequence of DDMs, and optimization to determine the decision thresholds that maximize the reward rate (RR). Due to the tractability of the DDM, the correct probability and decision times that constitute the RR can be computed analytically. We also demonstrate that response biases, such as repetitions, commonly observed in sequential decision tasks, follow naturally from this model. 

\section{Optimizing decisions in a sequence of uncorrelated trials}
\label{uncorr}

We first derive in~\ref{DDM_app} and summarize here the optimal evidence-accumulation model for a sequence of independent TAFC trials ($\P(H^{i} = H^{i-1}) = 0.5$). Although these results can be found in previous work~\citep{bogacz06},  they are crucial for the  subsequent discussion, and we thus provide them for completeness. \emph{A reader familiar with these classical results can skip to Section~\ref{corr}, and refer back to results in this section as needed.}

The drift-diffusion model (DDM) can be derived as the continuum limit of a recursive equation for the log-likelihood ratio (LLR)~\citep{wald48,gold02,bogacz06}. When the environmental states are uncorrelated and unbiased, an ideal observer has no bias at the start of each trial.

\subsection{Drift-diffusion model for a single trial}
\label{ddmsingle}
We assume that on a trial the observer integrates a stream of noisy measurements of the true state, $H^1$.  If these measurements
are conditionally independent, the functional central limit theorem yields the DDM for the scaled LLR,
$y^1(t) = D \log \frac{\P(H^1 = H_+  | \text{observations} )}{\P(H^1 = H_-  | \text{observations})},$ after  observation time $t$,
\begin{align}
\d y^1 &= g^1 \d t + \sqrt{2 D} \d W.  \label{ddm1}
\end{align}
Here $W$ is a Wiener process, the drift $g^1 \in g_{\pm}$ depends on the environmental state, and $D$ is the variance which depends on the noisiness of each observation. 

For simplicity and consistency with typical random dot kinetogram tasks~\citep{schall01,gold07}, we assume each drift direction is of equal unit strength: $g_+ = - g_- = 1$, and task difficulty is controlled by scaling the variance\footnote{An arbitrary drift amplitude $g$ can also be scaled out via a change of variables $y = g \tilde{y}$, so the resulting DDM for $\tilde{y}$ has unit drift, and $\tilde{D} = D/g^2$.} through $D$.  The initial condition is determined by the observer's prior bias, $y^1(0) = D \ln [ \P(H^1 = H_+) /  \P (H^1 = H_-) ]$.  Thus for an unbiased observer,  $y^1(0) = 0$.

There are two primary ways of obtaining a response from the DDM given by Eq.~(\ref{ddm1}) that mirror common experimental protocols~\citep{shadlen01,gold02,gold07}: An ideal  observer  interrogated at  a set time, $t = T$ responds with $\sign(y(T)) = \pm 1$ indicating the more likely of the two states, $H^1 = H_{\pm},$ given the accumulated evidence. On the other hand, an observer free to choose their response time can trade speed for accuracy in making a decision. This is typically modeled in the DDM by defining a decision threshold, $\theta_1,$ and assuming that at the first time, $T_1$, at which $|y^1(T_1)| \geq \theta_1$, the evidence accumulation process terminates, and the observer chooses $H_{\pm}$ if $\sign (y^1(T_1)) = \pm 1$. 

The probability, $c_1$, of making a correct choice in the free response paradigm can be obtained using the Fokker-Planck (FP) equation corresponding to Eq.~(\ref{ddm1})~\citep{gardiner09}. Given the initial condition $y^1(0) = y_0^1 = D \ln [ \P(H^1 = H_+)/\P(H^1= H_-)]$, threshold $\theta_1$, and  state $H^1 = H_+$, the probability of an exit through either boundary $\pm \theta_1$ is
\begin{align}
\pi_{\theta_1} (y_0^1) = \frac{1 - \e^{-(y_0^1 + \theta_1)/D}}{1 - \e^{- 2\theta_1/D}} , \hspace{10mm} \pi_{-\theta_1} (y_0^1) = \frac{\e^{- (y_0^1 + \theta_1)/D} - \e^{- 2\theta_1/D}}{1 - \e^{- 2\theta_1/D}},  \label{genpi}
\end{align}
simplifying at $y_0^1 = 0$ to
\begin{align}
\pi_{\theta_1} (0) = \frac{1}{1 + \e^{- \theta_1/D}} =: c_1, \hspace{12mm} \pi_{-\theta_1} (0) = \frac{\e^{- \theta_1/D}}{1 + \e^{- \theta_1/D}} = 1-c_1.   \label{zerpi}
\end{align}
An exit through the threshold $\theta_1$ results in a correct choice of $H^1 = H_+$, so $c_1 = \pi_{\theta_1}(0)$. The correct probability $c_1$ increases with $\theta_1$, since more evidence is required to reach a larger threshold. Defining the decision in trial 1 as $d_1 = \pm 1$ if $y^1(T_1) = \pm \theta_1$, Bayes' rule implies
\begin{subequations} \label{d1corprob}
\begin{align}
\frac{1}{1 + \e^{- \theta_1/D}} = c_1 = \P (d_1 = \pm 1 | H^1 = H_{\pm} ) =  \P (H^1 = H_{\pm} | d_1 = \pm 1), \\
\frac{\e^{- \theta_1/D}}{1 + \e^{-\theta_1/D}} = 1-c_1 = \P (d_1 = \pm 1 | H^1 = H_{\mp} ) = \P (H^1 = H_{\mp} | d_1 = \pm 1),
\end{align}
\end{subequations}
since $\P(H^1 = H_{\pm}) = \P(d_1 = \pm 1) = 1/2$. Rearranging the expressions in Eq.~(\ref{d1corprob}) and isolating $\theta_1$ relates the threshold $\theta_1$ to the LLR given a decision $d_1 = \pm 1$:
\begin{align}
\pm \theta_1 = D \ln \frac{\P (H^1 = H_+ | d_1 = \pm 1)}{\P (H^1 = H_- | d_1 = \pm 1)}.  \label{dec1toLR}
\end{align}



\subsection{Tuning performance via speed-accuracy tradeoff} Increasing $\theta_1$ increases the probability of a correct decision, and  the average time to make a decision, $DT_1$. Humans and other animals balance  speed and accuracy to maximize the rate of correct decisions~\citep{chittka09,bogacz10}. This is typically quantified using the reward rate (RR)~\citep{gold02}:
\begin{align}
\RR_1 (\theta_1) = \frac{c_1}{DT_1 + T_D}, \label{RR1}
\end{align}
where $c_1$ is the probability of a correct decision, $DT_1$ is the mean time required for $y^1(t)$ to reach either threshold $\pm \theta_1$, and $T_D$ is the prescribed time delay to the start of the next trial~\citep{gold02,bogacz06}. Eq.~(\ref{RR1}) increases with $c_1$ (accuracy) and with inverse time $1/(DT_1 + T_D)$ (speed). This usually leads to a nonmonotonic dependence of $\RR_1$ on the threshold, $\theta_1$, since increasing $\theta_1$ increases accuracy, but decreases speed~\citep{gold02,bogacz06}.

The average response time, $DT_1,$ can be obtained as the solution of a mean exit time problem for the Fokker--Planck (FP) equation for $p(y^1,t)$ with absorbing boundaries, $p(\pm \theta_1,t) = 0$~\citep{bogacz06,gardiner09}. Since the RR is determined by the average decision time over all trials, we compute the unconditional mean exit time,
\begin{align}
T(y_0^1; \theta_1) = \theta_1 \left[ \frac{\e^{\theta_1/D} + \e^{- \theta_1/D} - 2 \e^{-y_0^1/D}}{\e^{\theta_1/D} - \e^{- \theta_1/D}} \right] - y_0^1,  \label{Tygen}
\end{align}
which for $y_0^1 = 0$ simplifies to
\begin{align}
T(0; \theta_1) = \theta_1 \left[ \frac{1 - \e^{- \theta_1/D}}{1 + \e^{-\theta_1/D}} \right] = DT_1.  \label{DT1}
\end{align}
Plugging this expression into the RR as defined in Eq.~(\ref{RR1}), assuming $y_0^1 = 0$, we have 
\begin{align*}
\RR_1 (\theta_1) = \left(\theta_1 \left[ 1 - \e^{- \theta_1/D} \right] + T_D \left[ 1 + \e^{-\theta_1/D} \right] \right)^{-1}.
\end{align*}
We can identify the maximum of $\RR_1(\theta_1) > 0$ by finding the minimum of its reciprocal, $1/\RR_1(\theta_1)$~\citep{bogacz06}, 
\begin{align}
\theta_1^{\rm opt} = T_D + D - D {\mc W} \left( \exp \left[ (T_D+D)/D \right] \right).  \label{uncoropt}
\end{align}
Here ${\mc W}(f)$ is the Lambert W function (the inverse of $f({\mc W}) = {\mc W} \e^{\mc W}$). In the limit $T_D \to 0$, we have $\theta_1^{\rm opt} \to 0$, and Eq.~\eqref{uncoropt} defines nontrivial optimal thresholds at $T_D>0$.

Having established a policy for optimizing performance in a sequence of independent TAFC trials,  in the next section we consider dependent trials. We can again explicitly compute the RR function for sequential TAFC trials with states, $H^{1:n} = (H^1, H^2, H^3, ..., H^n)$, evolving according to a two-state Markov process with parameter $\ep : = \P (H^{n+1} \neq H^n)$. Unlike above, where $\ep = 0.5$, we will see that when $\ep \in [0, 0.5)$, an ideal observer starts each trial $H^n$ for $n \geq 2$ by using information obtained over the previous trial, to bias their initial belief, $y_0^n \neq 0$.

\section{Integrating information across two correlated trials}
\label{corr}


We first focus on the case of two sequential TAFC trials. Both states are equally likely at the first trial,  $\P(H^1 = H_{\pm}) = 1/2$, but the state at  the second trial can depend on the state of the first, $\ep : = \P(H^2 = H_{\mp}| H^1 = H_{\pm})$.   On both trials an observer makes observations, $\xi_s^n,$ with conditional density $f_{\pm}(\xi)$ if $H^n = H_{\pm}$  to infer the state $H^n$. The  observer also uses information from trial 1 to infer the state at trial 2. All information about $H^{1}$ can be obtained from the decision variable, $d_{1} = \pm 1$, and ideal observers use the decision variable to set their initial belief, $y_0^2 \neq 0,$ at the start of trial 2. We later show that the results for two trials can be extended to trial sequences of arbitrary length with states generated according to the same two-state Markov process.

\subsection{Optimal observer model for two correlated trials} 
\label{S:two.trials}

The first trial is equivalent to the single trial case discussed in section~\ref{ddmsingle}. Assuming each choice is equally likely, $\P(H^1 = H_{\pm}) = 1/2$, no prior information exists. Therefore, the decision variable, $y^1(t),$ satisfies the DDM given by Eq.~(\ref{ddm1}) with $y^1(0) = 0$. This generates a decision $d_1 \in \pm 1$ if $y^1(T_1) = \pm \theta_1$, so that the probability of a correct decision $c_1$ is related to the threshold $\theta_1$ by Eq.~(\ref{d1corprob}). Furthermore, since $\ep = \P ( H^2 = H_{\mp} | H^1 = H_{\pm})$, it follows that $1 - \ep = \P ( H^2 = H_{\pm} | H^1 = H_{\pm})$, which means
\begin{align}
\P (H^2 = H_{\pm} | d_1 = \pm 1) =& \P ( H^2 = H_{\pm} | H^1 = H_{\pm} ) \P (H^1 = H_{\pm} | d_1 = \pm 1) \nonumber \\
& + \P ( H^2 = H_{\pm} | H^1 = H_{\mp} ) \P (H^1 = H_{\mp} | d_1 = \pm 1) \nonumber \\
=& (1- \ep) c_1 + \ep (1-c_1),  \label{PH2same}
\end{align}
and, similarly,
\begin{align}
\P (H^2 = H_{\mp} | d_1 = \pm 1) &= (1- \ep) (1- c_1) + \ep c_1. \label{PH2diff}
\end{align}

As the decision, $d_1 = \pm 1,$ determines the probability of each state at the end of trial 1, individual observations, $\xi_{1:s}^1,$  are not needed to define the belief of the ideal observer at the outset of trial 2. The ideal observer uses a sequence of  observations, $\xi_{1:s}^2,$ and their decision on the previous trial, $d_1,$ to arrive at the probability ratio
\begin{align}
R_s^2 = \frac{\P (H^2 = H_+ | \xi_{1:s}^2, d_{1})}{\P (H^2 = H_- | \xi_{1:s}^2, d_{1})} = \frac{\P (\xi_{1:s}^2| H^2 = H_+) \P (H^2 = H_+ | d_{1})}{\P (\xi_{1:s}^2| H^2 = H_-) \P (H^2 = H_- | d_{1})}.  \label{LR2a}
\end{align}
Taking the logarithm, and applying conditional independence of the measurements $\xi_{1:s}^2$, we have that
\begin{align*}
L_s^2 = \ln R_s^2 = \sum_{j=1}^s \ln \frac{f_+(\xi_j^2)}{f_-(\xi_j^2)} + \ln \frac{\P (H^2 = H_+ | d_1)}{\P (H^2 = H_- | d_1)},
\end{align*}
indicating that $L_0^2 = \ln \left[ \P (H^2 = H_+ | d_1)/ \P (H^2 = H_- | d_1) \right]$. Taking the temporal continuum limit as in~\ref{DDM_app}, we find
\begin{align}
\d y^2 = g^2 \d t + \sqrt{2 D} \d W,  \label{ddm2}
\end{align}
with the Wiener process, $W$, and variance defined as in Eq.~(\ref{ddm1}). The drift $g^2 \in \pm 1$ is determined by $H^2 = H_{\pm}$.

Furthermore, the initial belief is biased in the direction of the previous decision, as the observer knows that states are correlated across trials. The initial condition for Eq.~\eqref{ddm2} is therefore
\begin{align}
y^2(0; d_1 = \pm 1) = D \ln \frac{\P (H^2 = H_+ | d_1 = \pm 1)}{\P (H^2 = H_- | d_1 = \pm 1)} = \pm D \ln \frac{(1 - \ep) \e^{\theta_1/D} + \ep}{\ep \e^{\theta_1/D} + (1- \ep)} = \pm y_0^2,  \label{y2ic}
\end{align}
where we have used Eqs.~(\ref{d1corprob}), (\ref{PH2same}), and (\ref{PH2diff}). Note that $y_0^2  \to 0$ in the limit $\ep \to 0.5$, so no information is carried forward when states are uncorrelated across trials. In the limit $\ep \to 0$, we find $y_0^2 \to \theta_1$, so the ending value of the decision variable $y^1(T_1) = \pm \theta_1$ in trial 1 is carried forward to trial 2, since there is no change in environmental state from trial 1 to 2. For $\ep \in (0,1/2)$, the information gathered on the previous trial provides partial information about the next state, and we have $y_0^2 \in (0, \theta_1)$.

We assume that the  decision variable, $y^2(t),$ evolves according to  Eq.~(\ref{ddm2}), until it reaches a threshold $ \pm \theta_2$, at which point the decision $d_2 = \pm 1$ is registered. The full model is thus specified by,
\begin{subequations} \label{twoDDM}
\begin{align}
\d y^j &= g^j \d t + \sqrt{2 D} \d W, \hspace{4mm} \text{until} \ |y^j(T_j)| \geq \theta_j \ \ \mapsto \ \  d_j = \sign(y^j(T_j)), \\
y^1(0) &= 0, \ \ \ \  y^2(0) = \pm y_0^2 \ \text{if} \ d_1 = \pm 1, \ \ \ \text{and} \ \ \ g^j = \pm 1 \ \text{if} \ H^j = H_{\pm}, j = 1,2. 
\end{align}
\end{subequations}

The above analysis is easily extended to the case of arbitrary $n \geq 2$, but before we do so, we analyze the impact of state correlations on the reward rate (RR) and identify the decision threshold values $\theta_{1:2}$ that maximize the RR.

{\bf Remark:} As noted earlier, we assume the total reward is given at the end of the experiment, and not after each
trial~\citep{Braun18}. A reward for a correct choice provides the observer with complete information about  the state $H^n$, so that
\begin{align*}
y^2(0; H^1 = \pm 1) = D \ln \frac{\P (H^2 = H_+ | H^1 = \pm 1)}{\P (H^2 = H_- | H^1 = \pm 1)} = \pm D \ln \frac{(1 - \ep) }{\ep}. 
\end{align*}
A similar result holds in the case that the reward is only given with some probability.

\subsection{Performance for constant decision thresholds} 

We next show  how varying the decision thresholds impacts the performance of the observer. For simplicity, we first assume the same threshold is used in both trials, $\theta_{1:2} = \theta$. We quantify performance using a reward rate (RR) function across both trials:
\begin{align}
\RR_{1:2} = \frac{c_1 + c_2}{DT_1 + DT_2 + 2 T_D},  \label{RR12}
\end{align}
where $c_j$ is the probability of correct responses on  trial $j = 1,2$, $DT_j$ is the mean decision time in trial $j$, and $T_D$ is the time delay after each trial. The expected time until the reward is thus $DT_1 + DT_2 + 2 T_D$. The reward rate will increase as the correct probability in each trial increases, and as decision time decreases. However, the decision time in trial 2 will depend on the outcome of trial 1. 

Similar to our finding for trial 1, the decision variable $y^2(t)$ in trial 2 determines  ${\rm LLR}_2(t),$ since  $y^2 = D \cdot {\rm LLR}_2$. Moreover, the probability of a correct responses on each trial, $c_1$ and $c_2$ are \emph{equal}, and determined by the threshold, $\theta = \theta_{1:2},$ and  noise amplitude, $D$ (See~\ref{c2thresh}).

\begin{figure}[t]
\begin{center} \includegraphics[width=12cm]{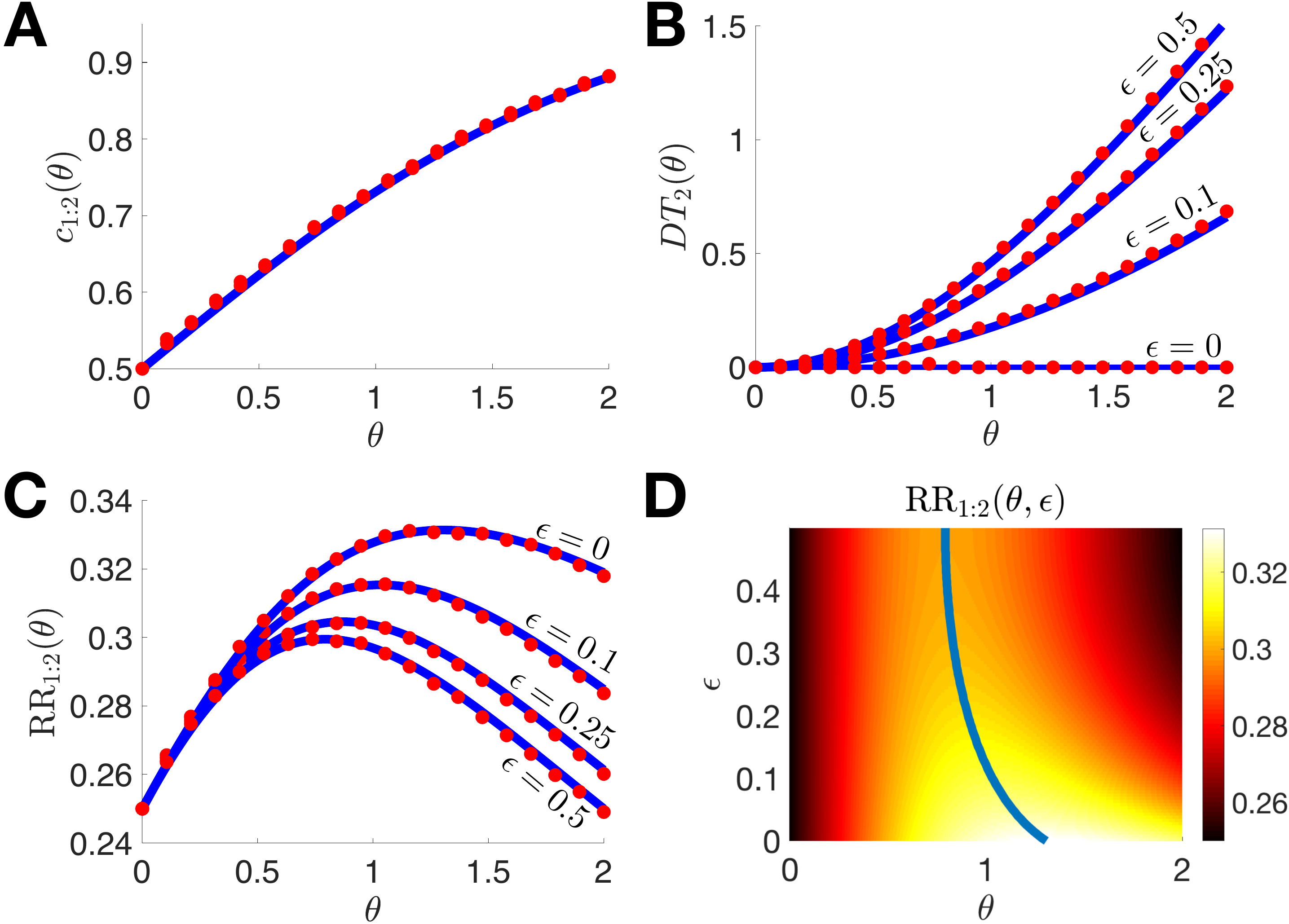} \end{center}
\caption{The reward rate (RR)  depends on the state change rate $\ep : = \P (H^2 \neq H^1)$ for a constant threshold $\theta : = \theta_{1:2}$. Here and below stochastic simulations (red dots) are compared to theoretical predictions (solid blue). {\bf A}. The  probability of a correct response in trials 1 and 2 ($c_{1:2}$) only depends on the threshold $\theta$, and increases with $\theta$.  {\bf B}. Decision time $DT_2$ in trial 2 increases with the threshold $\theta$ and with the change rate, $\ep$. Threshold crossings happen sooner on average for lower $\ep$, since trial 2 is initiated closer to threshold. 
{\bf C}. The reward rate (RR) as defined by Eq.~(\ref{RR12}) is unimodal. {\bf D}. Colormap plot of  RR as a function of $\theta$ and $\ep$. The maximal RR increases and occurs at higher thresholds, $\theta,$ when state changes are less likely (lower $\ep$): As $\ep$ decreases, the observer can afford to integrate evidence for a longer period of time in trial 1, since trial 2 will be much shorter (as shown in {\bf B}). Here and in subsequent figures we used delay time $T_D = 2$ and noise amplitude $D=1$. }
\label{fig2_twoconst}
\end{figure}

The average time until a decision in trial 2 is (See~\ref{avgtime})
\begin{align}
DT_2 = \frac{1-  \e^{- \theta/D}}{1+ \e^{- \theta/D}}\left[ \theta - (1- 2 \ep) D \ln \frac{(1- \ep) \e^{\theta/D} + \ep}{\ep \e^{\theta/D} + (1- \ep)} \right].  \label{DT2}
\end{align}
Notice, in the limit $\ep \to 0.5$, Eq.~(\ref{DT2}) reduces to Eq.~(\ref{DT1}) for $\theta_1 = \theta$ as expected. Furthermore, in the limit $\ep \to 0$, $DT_2 \to 0$, since decisions are made immediately in an unchanging environment as $y_0^2(0) = \pm \theta$.

We can now use the fact that $c_1 = c_2$ to combine Eqs.~\eqref{d1corprob}, and \eqref{DT2} with Eq.~(\ref{RR12}) to find
\begin{align*}
\RR_{1:2}(\theta)  &= 2 \left( \D \left[ 1 - \e^{- \theta/D} \right] \left[ 2 \theta - (1- 2\ep)  D \ln \frac{(1 - \ep) \e^{\theta_1/D} + \ep}{\ep \e^{\theta_1/D} + (1- \ep)} \right] + 2 T_D \left[ 1 + \e^{- \theta/D} \right] \right)^{-1},
\end{align*}
which can be maximized using numerical optimization.

Our analysis shows that the correct probability in both trials, $c_{1:2}$, increases with $\theta$ and does not depend on the transition probability $\ep$ (Fig. \ref{fig2_twoconst}A). In addition, the decision time in trial 2, $DT_2$, increases with $\theta$ and $\ep$ (Fig. \ref{fig2_twoconst}B). At smaller transition rates, $\ep,$  more information is carried forward to the second trial, so the decision variable $y_0^2$ is, on average, closer to the threshold $\theta$. This shortens the average decision time, and increases the  RR (Fig. \ref{fig2_twoconst}C, D).  Lastly, note that the threshold value, $\theta,$ that maximizes the RR is higher for lower values of $\ep$, as the observer can afford to require more evidence for a decision when starting trial 2 with more information.

\subsection{Performance for dynamic decision thresholds} 

We next ask whether the RR given by Eq.~(\ref{RR12}) can be increased by allowing unequal decision thresholds between the trials,  $\theta_1 \neq \theta_2$. As before, the  probability of a correct response, $c_1 = \pi_{\theta_1}(0),$ and mean decision time, $DT_1,$ in trial 1 are given by Eqs.~(\ref{zerpi}) and (\ref{DT1}). The correct probability $c_2$ in trial 2 is again determined by the decision threshold, and noise variance, $D$,
 as long as $\theta_2 > y_0^2$. However, when $\theta_2< y_0^2$, the response in trial 2 is instantaneous ($DT_2 = 0$) and $c_2 = (1- \ep) c_1 + \ep (1-c_1)$. Therefore, the probability of a correct response on trial 2 is defined piecewise
\begin{align}
c_2 &= \left\{ \begin{array}{cc} \D  \frac{1}{1 + \e^{- \theta_2/D}} & : \ \theta_2 > y_0^2, \\  \D \frac{(1 - \ep) + \ep \e^{-\theta_1/D}}{1 + \e^{-\theta_1/D}} & : \theta_2 \leq y_0^2. \end{array} \right.  \label{c2dyn}
\end{align}
We will show in the next section that this result extends to an arbitrary number of trials, with an arbitrary sequence of thresholds, $\theta_{1:n}$.

\begin{figure}[t]
\begin{center} \includegraphics[width=12cm]{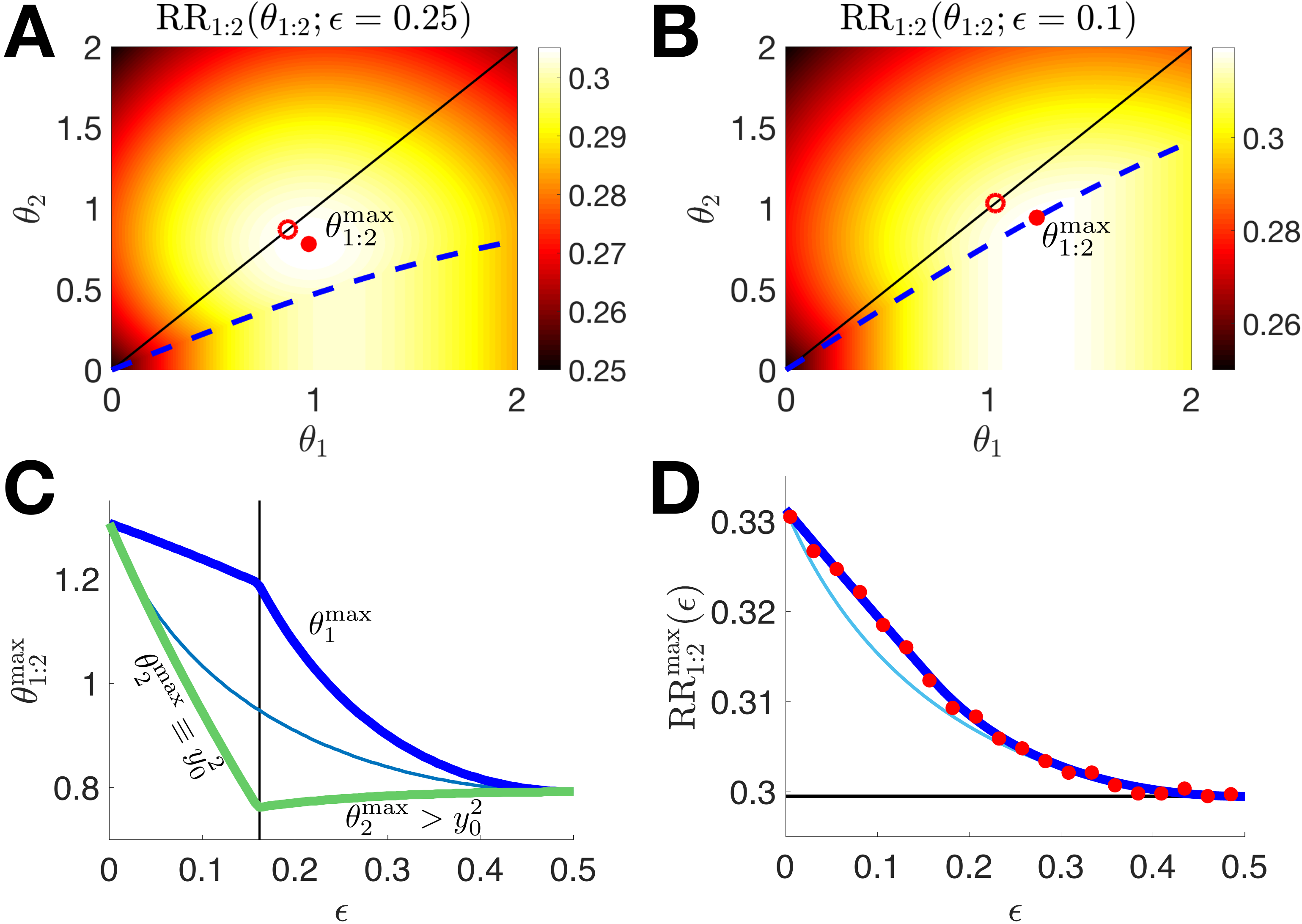} \end{center}
\caption{{\bf A},{\bf B}. Reward rate (RR), given by Eq.~(\ref{RRdyn}), depends on the change rate $\ep$ and thresholds $\theta_{1:2}$. For $\ep = 0.25$ (panel {\bf A}), the pair $\theta_{1:2}^{\rm max}$ that maximizes  RR (red dot) satisfies $\theta_1^{\rm max}> \theta_2^{\rm max}$. Compare to the  pair $\theta_{1:2}^{\rm max}$ (red circle) for $\theta = \theta_{1:2}$ as in Fig. \ref{fig2_twoconst}D. The boundary $\theta_2 = y_0^2(\theta_1)$, obtained from Eq.~(\ref{y2ic}), is shown as a dashed blue line. For $\ep = 0.1$ (panel {\bf B}), the optimal choice of thresholds (red dot) is on the boundary $\theta_2 = y_0^2(\theta_1)$. {\bf C}. Decreasing $\ep$ from 0.5,  initially results in an increase in $\theta_1^{\rm max},$ and decrease in  $\theta_2^{\rm max}$. At sufficiently small value of $\ep$, the second threshold collides with the initial condition, $\theta_2 = y_0^2(\theta_1)$  (vertical black line). After this point, both $\theta_{1}^{\rm max}$ and $\theta_2^{\rm max}$ increase with $\ep$, and meet  at $\ep = 0$. The optimal constant threshold,  $\theta^{\rm max}$,  is represented by a thin blue line. {\bf D}.  $RR_{1,2}^{\rm max}(\ep)$ decreases as $\ep$ is increased, since the observer carries less information  from trial 1 to trial 2. 
The thin blue line shows  $RR_{1,2}^{\rm max}$  for the constant threshold case. 
}
\label{fig3_dynthresh}
\end{figure}

In the case $\theta_2 > y_0^2$, the average time until a decision in trial 2, is  (See~\ref{avgtime})
\begin{equation}
DT_2 
= \theta_2 \frac{1 -\e^{-\theta_2/D}}{1+\e^{-\theta_2/D}} - \frac{(1- 2\ep) (1 - \e^{- \theta_1/D})}{1+\e^{-\theta_1/D}} D \ln \frac{(1- \ep) \e^{\theta_1/D} + \ep}{\ep \e^{\theta_1/D} + (1- \ep)} , \label{E:avgdyn2}
\end{equation}
which allows us to compute the reward rate function ${\rm RR}_{1:2}(\ep) = {\rm RR}_{1:2}(\theta_{1:2}; \ep)$:
\begin{align}
{\rm RR}_{1:2}(\ep) = \left\{ \begin{array}{cc} \D \frac{E_+^1 + E_+^2 }{E_+^2 E_-^1 \left[ \theta_1 - (1-2\ep) y_0^2 \right]  + E_+^1 \left[ \theta_2 E_-^2 + 2T_D E_+^2 \right] } & : \ \theta_2 > y_0^2, \\[\bigskipamount]
\D \frac{2(1 - \ep) + \ep E_+^1}{\theta_1 E_-^1 + 2 T_D E_+^1} & : \ \theta_2 \leq y_0^2, \end{array} \right.  \label{RRdyn}
\end{align}
where $E_{\pm}^j = 1 \pm \e^{- \theta_j/D}$. This reward rate is convex as a function of the thresholds, $\theta_{1:2},$ for all examples we examined.

The pair of thresholds, $\theta_{1:2}^{\rm max},$ that maximize RR satisfy $\theta_1^{\rm max} \geq \theta_2^{\rm max}$, so that the optimal observer typically decreases their decision threshold from trial 1 to trial 2 (Fig. \ref{fig3_dynthresh}A,B). This shortens the mean time to a decision in trial 2,  since the initial belief $y_0^2$ is closer to the decision threshold more likely to be correct.
As the change probability $\ep$  decreases from 0.5, initially the optimal $\theta_2^{\rm max}$ decreases and $\theta_1^{\rm max}$ increases (Fig. \ref{fig3_dynthresh}C), and $\theta_2^{\rm max}$ eventually colides with the boundary $y_0^2 = \theta_2$. For smaller  values of $\ep$, RR is  thus maximized when the observer accumulates information on trial 1, and then makes the same decision instantaneously on trial 2, so that $c_2 = (1- \ep) c_1 + \ep (1-c_1)$ and $DT_2 = 0$. As in the case of a fixed threshold, the maximal reward rate, ${\rm RR}_{1,2}^{\rm max}(\ep),$ increases as $\ep$ is decreased (Fig. \ref{fig3_dynthresh}D), since the observer starts trial 2 with more information. Surprisingly, despite different strategies,  there is no big increase in ${\rm RR}_{1,2}^{\rm max}$ compared to constant thresholds, with the largest gain at low values of $\ep$. \\
\vspace{-4mm}

{\bf Remark:} In the results displayed for our RR maximization analysis (Fig. \ref{fig2_twoconst} and \ref{fig3_dynthresh}), we have fixed $T_D =2$ and $D=1$. We have also examined these trends in the RR at lower and higher values of $T_D$ and $D$ (plots not shown), and the results are qualitatively similar. The impact of dynamic thresholds $\theta_{1:2}$ on the RR are slightly more pronounced when $T_D$ is small or when $D$ is large.

\section{Multiple correlated trials}
\label{ntrials}

To understand optimal decisions across multiple correlated trials, we again assume that the states evolve according to a two-state Markov Chain with  $\ep : = \P (H^{j+1} = H_{\mp} | H^j = H_{\pm})$. Information gathered on one trial again determines the initial belief on the next. We first discuss optimizing the RR for  a constant threshold for a fixed number $n$ of trials known to the observer.  When decision thresholds are allowed to vary between trials,  optimal thresholds typically decrease across trials, $\theta_j^{\rm max} \leq \theta_{j-1}^{\rm max}$. 

\subsection{Optimal observer model for multiple trials} As in the case of two trials, we must consider the caveats associated with instantaneous decisions in order to define initial conditions $y_0^j$, probabilities of correct responses $c_j$, and mean decision times $DT_j$. The same argument  leading to Eq.~\eqref{ddm2} shows that the decision variable evolves according to a linear drift diffusion process on each trial.
The probability of a correct choice is again defined by the threshold, $c_j = 1/(1 + \e^{- \theta_j/D})$  when $\theta_j > y_0^j$, and $c_j = (1- \ep) c_{j-1} + \ep (1 - c_{j-1})$ when $\theta_j \leq y_0^j$,  as in Eq.~(\ref{c2dyn}). Therefore, $c_j$ may be defined iteratively for $j \geq 2$, as 
\begin{align}
c_j &= \left\{ \begin{array}{cc} \D \frac{1}{1 + \e^{- \theta_j/D}} & : \ \theta_j > y_0^j, \\
(1- \ep) c_{j-1} + \ep (1 - c_{j-1}) & : \ \theta_j \leq y_0^j, \end{array} \right.  \label{cj}
\end{align}
with $c_1 = 1/(1 + \e^{- \theta_1/D})$  since $y_0^1= 0$.

The probability $c_j$  quantifies the belief  at the end of trial $j$. Hence,  
as in Eq.~\eqref{y2ic}, the initial belief of an ideal observer on trial $j+1$ has the form
\begin{align}
y^{j+1}(0; d_j = \pm 1) = \pm D \ln \frac{(1- \ep) c_j + \ep (1-c_j)}{\ep c_j + (1-\ep) (1-c_j) } = \pm y_0^{j+1},  \label{yjic}
\end{align}
and the decision variable $y^j(t)$ within trials obeys a DDM with threshold $\theta_j$ as in Eq.~(\ref{ddm2}).
A decision $d_j = \pm 1$ is then registered when $y^j(T_j) = \pm \theta_j$, and if $\theta_j \leq y_0^j$, then $d_j = d_{j-1}$ and $T_j = 0$.

\subsection{Performance for constant decision thresholds} With $n$ trials,
the RR is the sum of correct probabilities divided by the total decision and delay time:
\begin{align}
\RR_{1:n} = \frac{ \sum_{j=1}^n c_j}{\sum_{j=1}^n DT_j + n T_D}.  \label{RRn}
\end{align}
The  probability of a correct choice, $c_j = 1/(1 + \e^{- \theta/D})$,  is constant across trials, and  determined by the  fixed threshold $\theta$. It follows that $y_0^2 = y_0^3 = \cdots = y_0^n = y_0$. Eq.~(\ref{yjic}) then implies that $\theta > y_0^j$ for $\ep >0$ and $y_0^j = \theta$ if $\ep = 0$.  

\begin{figure}
\begin{center} \includegraphics[width=16cm]{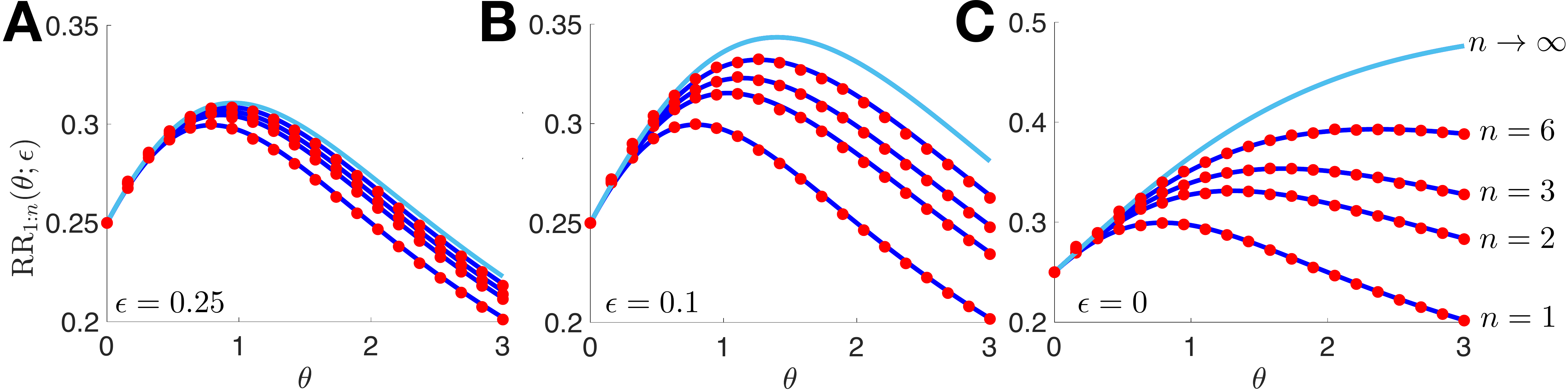} \end{center}
\caption{Reward rate (${\rm RR}_{1:n}$) defined by Eq.~(\ref{RRconstn}) increases with $n$, and as $\ep$ is decreased, as shown for $\ep = 0.25$ (panel {\bf A}); $\ep = 0.1$ (panel {\bf B}); and $\ep = 0$ (panel {\bf C}). The optimal decision threshold, $\theta^{\rm max},$ increases with $n$. 
The limiting reward rates ${\rm RR}_{1:n}(\theta;\ep) \to {\rm RR}_{\infty}(\theta; \ep)$ as $n \to \infty$ defined by Eq.~(\ref{RRconstinf}) are shown in light blue. }
\label{fig4_multconst}
\end{figure}

As $y_0^j$ is constant on the 2nd trial and beyond, the  analysis of the two trial case implies that the mean decision time is given by an expression equivalent to Eq.~(\ref{DT2}) for $j \geq 2$.
Using these expressions for $c_j$ and $DT_j$ in Eq.~(\ref{RRn}), we find
\begin{align}
{\rm RR}_{1:n} (\theta; \ep) = n \left( \left[ 1 - \e^{- \theta/D} \right] \left[ n \theta - (n-1)(1-2 \ep) y_0 \right] + n T_D \left[ 1 + \e^{- \theta/D} \right] \right)^{-1}.  \label{RRconstn}
\end{align}
As $n$ is increased, the threshold value, $\theta^{\rm max},$ that maximizes the RR increases  across the full range of $\ep$ values (Fig. \ref{fig4_multconst}): As the number of trials increases, trial 1  has less impact on the RR.  Longer decision times on trial 1 impact the RR less, allowing for longer accumulation times (higher thresholds) in later trials. In the limit of an infinite number of trials ($n \to \infty$), Eq.~(\ref{RRconstn}) simplifies to
\begin{align}
{\rm RR}_{\infty} (\theta; \ep)  =  \left( \left[ 1 - \e^{- \theta/D} \right] \left[ \theta - (1-2 \ep) y_0 \right] + T_D \left[ 1 + \e^{- \theta/D} \right]\right)^{-1} ,  \label{RRconstinf}
\end{align}
and we find $\theta^{\rm max}$ of ${\rm RR}_{\infty}(\theta; \ep)$ sets an upper limit on the optimal $\theta$ for $n< \infty$. 

Interestingly, in a static environment ($\ep \to 0$), 
\begin{align}
\lim_{\ep \to 0} {\rm RR}_{\infty}(\theta; \ep) = \frac{1}{T_D \left[ 1 + \e^{- \theta/D} \right]},
\end{align}
so the decision threshold value that maximizes RR diverges, $\theta^{\rm max} \to \infty$ (Fig. \ref{fig4_multconst}C). Intuitively, when there are many trials ($n \gg 1$), the price of a long wait for a high accuracy decision in the first trial ($c_1 \approx 1$ but $DT_1 \gg 1$ for $\theta \gg 1$) is offset by the reward from a large number ($n-1 \gg 1$) of subsequent, instantaneous, high accuracy decisions ($c_j \approx 1$ but $DT_2 =0$ for $\theta \gg 1$).

\subsection{Dynamic decision thresholds}
\label{S:dynamic.multi}

 In the case of dynamic thresholds, the RR function, Eq.~(\ref{RRn}), can be maximized for an arbitrary number of trials, $n$. The probability of a correct decision, $c_j,$ is given by the more general, iterative version of Eq.~(\ref{cj}). Therefore, while analytical results can be obtained for constant thresholds, numerical methods are needed to find the sequence of optimal dynamic thresholds.

The expression for the mean decision time, $DT_j,$ on trial $j$ is determined by marginalizing over whether or not the initial state $y_0^j$ aligns with the true state $H^j$, yielding
\begin{align}
DT_j &= \left\{ \begin{array}{cc}  \D \theta_j \frac{1 - \e^{- \theta_j/D}}{1 + \e^{- \theta_j/D}} - (1 - 2 \ep) (2c_{j-1} -1) y_0^j & : \ \theta_j > y_0^j, \\
0 & : \ \theta_j \leq y_0^j. \end{array} \right.  \label{DTj}
\end{align}
Thus,  Eq.~(\ref{cj}), the probability of a correct decision on trial $j-1$, is needed to determine the mean decision time, $DT_j,$ on trial $j$. The resulting values for $c_j$ and $DT_j$ can be used in Eq.~(\ref{RRn}) to determine the RR which again achieves a maximum for some choice of thresholds,  $\theta_{1:n}: = (\theta_1, \theta_2, ..., \theta_n)$.

\begin{figure}
\begin{center} \includegraphics[width=16cm]{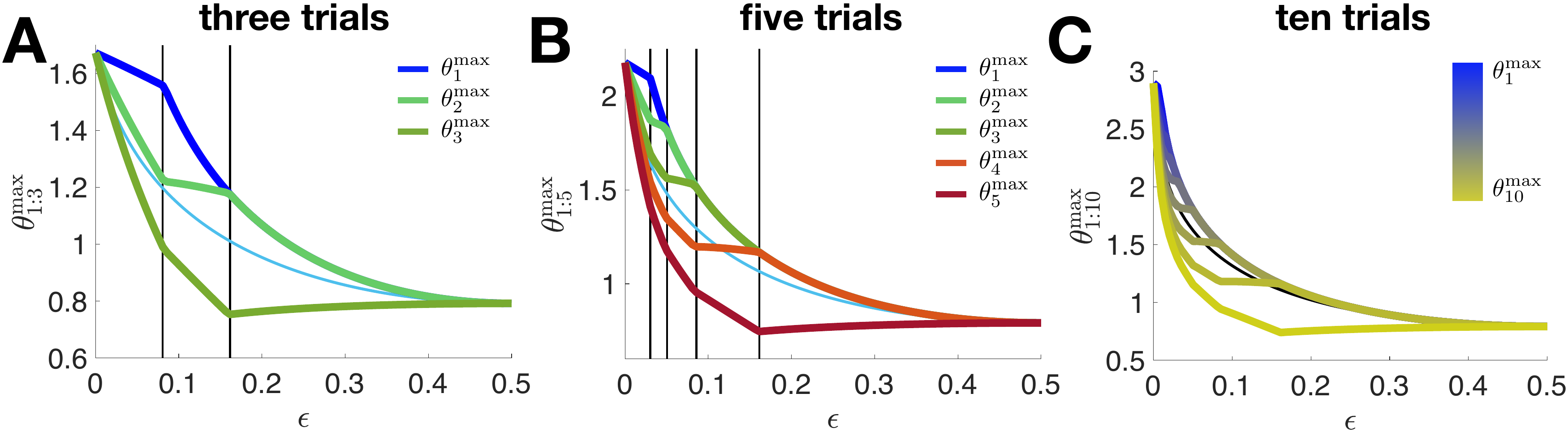} \end{center}
\caption{Optimal sets of thresholds, $\theta_{1:n}^{\rm max},$ for $n$ trials with correlated states. {\bf A}. For $n=3$, the set of optimal thresholds $\theta_{1:3}^{\rm max}$ obeys the ordering $\theta_1^{\rm max} \geq \theta_2^{\rm max} \geq \theta_3^{\rm max}$ for all values of $\ep$, converging as $\ep \to 0.5^-$ and as $\ep \to 0^+$. Vertical lines denote points below which $\theta_j^{\rm max} = y_0^j$ successively for $j=3,2,1$. {\bf B},{\bf C}. Ordering of optimal thresholds is consistent in the case of $n=5$ (panel {\bf B}) and $n=10$ trials (panel {\bf C}). 
}
\label{fig5_multdyn}
\end{figure}

In Fig.~\ref{fig5_multdyn} we show that the sequence of decision thresholds that maximize RR,  $\theta_{1:n}^{\rm max}$, is decreasing ($\theta_1^{\rm max} \geq \theta_2^{\rm max} \geq \ldots \geq \theta_n^{\rm max}$) across trials, consistent with our observation for two trials. Again, the increased accuracy in earlier trials improves accuracy in later trials, so there is value in gathering more information early in the trial sequence.  Alternatively, an observer has less to gain from waiting for more information in a late trial, as the additional evidence can only be used on a few more trials.  For intermediate to large values of $\ep$, the observer uses a near constant threshold on the first $n-1$ trials, and then a lower threshold on the last trial.  As $\ep$ is decreased, the last threshold, $\theta_n^{\rm max}$, collides with the boundary $y_0^n$, so the last decision is made instantaneously.  The previous thresholds, $\theta_{n-1}^{\rm max}, \theta_{n-2}^{\rm max}, \ldots$ collide with the corresponding boundaries  $y_0^{n-1}, y_0^{n-2}, \ldots$ in reverse order as $\ep$ is decreased further. Thus for lower values of $\ep$, an increasing number of decisions toward the end of the sequence are made instantaneously.

\section{Comparison to experimental results on repetition bias}
\label{repbias}

We next discuss the experimental predictions of our model, motivated by previous observations of repetition bias. It has been long known that trends appear in the response sequences of subjects in series of TAFC trials~\citep{Fernberger20},  even when the trials are uncorrelated~\citep{cho02,gao09}. For instance, when the state is the same on two adjacent trials, the probability of a correct response on the second trial increases~\citep{Frund14}, and the response time decreases~\citep{jones13}. In our model,  if the assumed change rate of the environment does not match the true change rate, $\ep^{\rm assumed} \neq \ep^{\rm true}$, performance is decreased compared to $\ep^{\rm assumed} = \ep^{\rm true}$. However, this decrease may not be severe.  Thus the observed response trends may arise partly due to misestimation of the transition rate $\ep,$ and the assumption that trials are dependent, even if they are not~\citep{cho02,Yu08}.

Motivated by our findings on ${\rm RR}_{1:n}^{\rm max}$ in the previous sections, we focus on the model of an observer who fixes their decision threshold across all trials, $\theta_j = \theta$, $\forall j$. Since allowing for dynamic thresholds does not impact ${\rm RR}_{1:n}^{\rm max}$ considerably, we think the qualitative trends we observe here should generalize.

\subsection{History-dependent psychometric functions} 

\begin{figure}
\begin{center} \includegraphics[width=12cm]{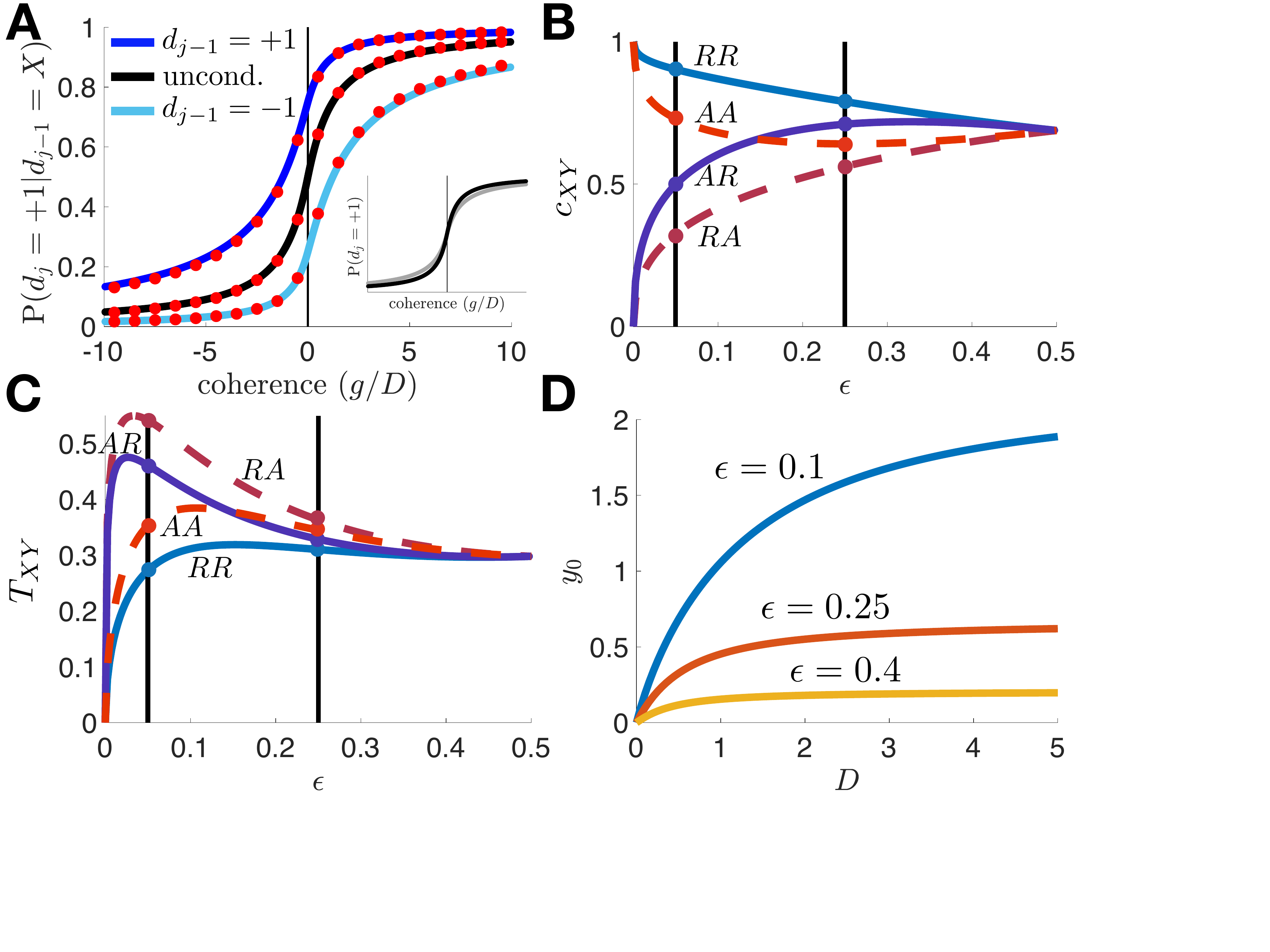} \end{center}
\caption{Repetition bias trends in the adaptive DDM. {\bf A}. The psychometric functions in trial $j$ conditioned on $d_{j-1}$ shows that when $d_{j-1} = +1$, the fraction of $d_j = +1$ responses increases (dark blue). When $d_{j-1} = -1$, the fraction of $d_j = +1$ responses decreases (light blue). The weighted average of these two curves gives the unconditioned psychometric function (black curve). Coherence $g/D$ is defined for drift $g \in \pm 1$, noise diffusion coefficient $D$, and assumed change rate $\ep = 0.25$.   Inset: When $\ep^{\rm true} = 0.5$ is the true change rate, the psychometric function is shallower (grey curve). {\bf B}. Correct probability $c_{XY}$ in current trial, plotted as a function of the assumed change rate $\ep$, conditioned on the pairwise relations of the last three trials ($R$: repetitions; $A$: alternations). Note the ordering of $c_{AA}$ and $c_{AR}$ is exchanged as $\ep$ is increased, since the effects of the state two trials back are weakened. Compare colored dots at $\ep=0.05$ and $\ep=0.25$. {\bf C}. Decision time $T_{XY}$ conditioned on pairwise relations of last three trials. {\bf D}. The initial condition, $y_0,$ increases as a function of the noise diffusion coefficient $D$ for $\ep = 0.1, 0.25, 0.4$. In all cases, we set $T_D = 2$, $\theta = \theta_{\infty}^{\rm max}$ for all trials, and $D=1$ if not mentioned.}
\label{fig6_compare}
\end{figure}

To provide a normative explanation of repetition bias, we  compute the degree to which the optimal observer's choice on  trial $j-1$ biases their choice on trial $j$. For comparison with experimental work, we  present the probability that $d_j = +1$, conditioned both on the choice on the previous trial, and on coherence\footnote{We define coherence as the drift, $g^j \in \pm1,$ in trial $j$ divided by the noise diffusion coefficient $D$~\citep{gold02,bogacz06}.} (See Fig.~\ref{fig6_compare}A) . In this case, $\P (d_j = +1 | g^j/D, d_{j-1} = \pm 1) = \pi_{\theta} (\pm y_0^j)$, where the exit probability $\pi_{\theta} (\cdot)$ is defined by Eq.~(\ref{genpi}), and the initial decision variable, $y_0^j,$ is given by Eq.~\eqref{y2ic} using the threshold $\theta$ and the assumed transition rate, $\ep = \ep^{\rm assumed}$.
The unconditioned psychometric function is obtained by marginalizing over the true environmental change rate, $\ep^{\rm true}$, $\P (d_j = +1 | g^j/D) = (1- \ep^{\rm true}) \P (d_j = +1 | g^j/D, d_{j-1} = + 1) + \ep^{\rm true}  \P (d_j = +1 | g^j/D, d_{j-1} = - 1)$, which equals $\pi_{\theta}(0)$ if $\ep = \ep^{\rm true}$.

Fig. \ref{fig6_compare}A shows that the probability of decision $d_{j} = +1$ is increased (decreased) if the same (opposite) decision was made on the previous trial.  When $\ep^{\rm assumed} \neq \ep^{\rm true}$, the psychometric function is shallower and overall performance worsens (inset), although only slightly. Note that we have set $\theta = \theta_{\infty}^{\rm max}$, so the decision threshold  maximizes $\RR_{\infty}$ in Eq.~(\ref{RRconstinf}). The increased probability of a repetition arises from the fact that the initial condition $y^j(0)$ is biased in the direction of the previous decision.

\subsection{Error rates and decision times} 

The adaptive DDM also exhibits repetition bias in decision times~\citep{goldfarb12}.
If the state at the current trial is the same as that on the previous trial (repetition) the decision time is shorter than average, while the decision time is increased if the state is different (alternation, see~\ref{rep_app}). This occurs for optimal ($\ep^{\rm assumed} = \ep^{\rm true}$) and suboptimal ($\ep^{\rm assumed} \neq \ep^{\rm true}$) models as long as $\ep \in [0,0.5)$. Furthermore, the rate of correct responses increases for repetitions and decrease for alternations (See~\ref{rep_app}). However,  the impact of the state on two or more previous trials on the choice in the present trial becomes more complicated. As in previous studies~\citep{cho02,goldfarb12,jones13}, we propose that such repetition biases in the case of uncorrelated trials could  be the result of the subject assuming $\ep = \ep^{\rm assumed} \neq 0.5$, even when 
$\ep^{\rm true} = 0.5$.

Following \citet{goldfarb12}, we denote that three adjacent trial states as composed of repetitions ($R$), and alternations ($A$)~\citep{cho02,goldfarb12}. Repetition-repetition ($RR$)\footnote{Note that $RR$ here refers to repetition-repetition, as opposed to earlier when we used ${\rm RR}$ in Roman font to denote reward rate.} corresponds to the three identical  states in a sequence, $H^j = H^{j-1} = H^{j-2}$; $RA$ corresponds to $H^j \neq H^{j-1} = H^{j-2}$; $AR$ to $H^j = H^{j-1} \neq H^{j-2}$; and $AA$ to $H^j \neq H^{j-1} \neq H^{j-2}$. The correct probabilities $c_{XY}$ associated with all four of these cases can be computed explicitly, as shown in \ref{rep_app}. Here we assume that $\ep^{\rm assumed} = \ep^{\rm true}$ for simplicity, although it is likely that subject often use $\ep^{\rm assumed} \neq \ep^{\rm true}$ as shown in \citet{cho02,goldfarb12}.

The probability of a correct response as a function of $\ep$ is shown for all four conditions in Fig. \ref{fig6_compare}B. Clearly, $c_{RR}$ is always largest, since the ideal observer correctly assumes repetitions are more likely than alternations (when $\ep \in [0,0.5)$). 

Note when $\ep = 0.05$, $c_{RR} > c_{AA}> c_{AR} > c_{RA}$ (colored dots), so a sequence of two alternations yields a higher correct probability on the current trial than a repetition preceded by an alternation. This may seem counterintuitive, since we might expect repetitions to always yield higher correct probabilities. However, two alternations yield the same state on the first and last trial in the three trial sequence ($H^j = H^{j-2}$). For small $\ep$, since initial conditions begin close to threshold, then it is likely $d_j = d_{j-1}=d_{j-2}$, so if $d_{j-2}$ is correct, then $d_j$ will be too.  At small $\ep$, and at parameters we chose here, the increased likelihood of the first and last response being correct, outweighs the cost of the middle response likely being incorrect.
Note that $AA$ occurs with probability $\ep^2$, and is thus rarely observed  when $\ep$ is small. As $\ep$ is increased, the ordering switches to $c_{RR} > c_{AR}> c_{AA} > c_{RA}$ (e.g., at $\ep = 0.25$), so a repetition in trial $j$ leads to higher probability of a correct response. In this case, initial conditions are less biased, so the effects of the state two trials back are weaker. As before, we have fixed $\theta = \theta_{\infty}^{\rm max}$.

These trends extend to decision times. Following a similar calculation as used for correct response probabilities, $c_{XY}$, we can compute the mean decision times, $T_{XY}$, conditioned on all possible three state sequences (See \ref{rep_app}). Fig. \ref{fig6_compare}C shows that, as a function of $\ep$, decision times following two repetitions, $T_{RR}$, are always the smallest. This is because  the initial condition in trial $j$ is most likely to start closest to the correct threshold: Decisions do not take as long when the decision variable starts close to the boundary. Also, note that when $\ep = 0.05$, $T_{RR}< T_{AA}< T_{AR} < T_{RA}$, and the explanation is similar to that for the probabilities of correct responses. When the environment changes slowly, decisions after two alternations will be quicker because a bias in the direction of the correct decision is more likely. 
The ordering is different at intermediate $\ep$ where $T_{RR} < T_{AR}< T_{AA} < T_{RA}$.

Notably, both orderings of the correct probability and decision times, as they depend on the three trial history, have been observed in previous studies of serial bias~\citep{cho02,gao09,goldfarb12}: Sometimes $c_{AA}>c_{AR}$ and $T_{AA}<T_{AR}$ whereas sometimes $c_{AA}<c_{AR}$ and $T_{AA}>T_{AR}$. However, it appears to be more common that $AA$ sequences lead to lower decision times, $T_{AA}<T_{AR}$ (e.g., See Fig.~3 in \citet{goldfarb12}). This suggests that subjects assume state switches are rare, corresponding to a smaller $\ep^{\rm assumed}$.

\subsection{Initial condition dependence on noise} Lastly, we explored how the initial condition, $y_0$, which captures the information gathered on the previous trial, changes with signal coherence, $g/D$.
We only consider the case  $\theta = \theta_{\infty}^{\rm max}$, and note that the threshold will vary with $D$. We find that the initial bias, $y_0$, always tends to increase with $D$ (Fig. \ref{fig6_compare}D). The intuition for this result is that the optimal $\theta = \theta_{\infty}^{\rm max}$ will tend to increase with $D$, since uncertainty increases with $D$, necessitating a higher threshold to maintain constant accuracy (See Eq.~(\ref{d1corprob})). As the optimal threshold increases with $D$, the initial bias increases with it. This is consistent with experimental observations that show that bias tends to decrease with coherence, \emph{i.e.} increase with noise amplitude~\citep{Frund14,Olianezhad16,Braun18}.

We thus found that experimentally observed history-dependent perceptual biases can be explained using a model of an ideal observer accumulating evidence across correlated TAFC trials. Biases are stronger when the observer assumes a more slowly changing environment -- smaller $\ep$ -- where trials can have an impact on the belief of the ideal observer far into the future. Several previous studies have proposed this idea~\citep{cho02,goldfarb12,Braun18}, but to our knowledge, none have used a normative model to explain and quantify these effects.

\section{Discussion}
\label{discussion}

It has been known for nearly a century that observers take into account previous choices when
making decisions~\citep{Fernberger20}.    While a number of descriptive models have
been used to explain experimental data~\citep{Frund14,goldfarb12}, there have been no normative models
that quantify how information accumulated on one trial impacts future decisions.  Here we have shown that a straightforward, tractable extension of previous evidence-accumulation models can be used to do so.

To account for information obtained on one trial  observers could adjust accumulation speed~\citep{ratcliff85,diederich06,urai18}, threshold~\citep{bogacz06,goldfarb12,diederich06}, or their initial belief on the subsequent trial~\citep{bogacz06,Braun18}.   We have shown that an ideal observer adjusts their initial belief and decision threshold, but not the accumulation speed in a sequence of dependent, statistically-identical trials. Adjusting the initial belief increases reward rate by decreasing response time, but not the probability of a correct response. Changes in initial bias have a larger effect than adjustments in threshold, and there is evidence that human subjects do adjust their initial beliefs across trials~\citep{Frund14,Olianezhad16,Braun18}. On the other hand, the effect of a dynamic threshold across trials diminishes as the number of trials increases: ${\rm RR}_{1:n}^{\rm max}$ is nearly the same for constant thresholds $\theta$ as for when allowing dynamic thresholds $\theta_{1:n}$. Given the cognitive cost of trial-to-trial threshold adjustments, it is thus unclear whether human subjects would implement such a strategy~\citep{Balci11}.

To isolate the effect of previously gathered information, we have assumed that 
the observers do not receive feedback or reward until the end of the trial. Observers can thus
use only information gathered on previous trials to adjust their bias on the next.  However, our results can be easily extended to the case when feedback is provided on each trial. The expression for the initial condition $y_0$ is simpler in this case, since feedback provides certainty concerning the state on the previous trial.

We also assumed that an observer knows the length of the trial sequence, and uses this number to optimize reward rate. It is rare that an organism would have such information in a natural situation.  However, our model can be extended to random sequence durations if we allow observers to marginalize over possible sequence lengths. If decision thresholds are fixed to be constant across all trials $\theta_j = \theta$, the projected RR is the average number  of correct decisions $c \langle n \rangle$ divided by the average time of the sequence. For instance, if the trial length follows a geometric distribution, $n \sim {\rm Geom}(p),$ then  
\begin{align*}
\bar{\RR}(\theta; \ep)  = \frac{1/p}{ \left[ 1 - \e^{- \theta/D} \right] \left[ \theta/p - (1/p-1)(1-2 \ep) y_0 \right] + (T_D/p) \left[ 1 + \e^{- \theta/D} \right]}.
\end{align*}
This equation has the same form as Eq.~(\ref{RRconstn}), the RR in the case of fixed trial number $n$, with $n$ replaced by the average $\langle n \rangle = 1/p$. Thus, we expect that as $p$ decreases, and the average trial number, $1/p,$ increases, resulting in an increase in the RR and larger optimal decision thresholds $\theta^{\rm max}$ (as in Fig. \ref{fig4_multconst}). The case of
dynamic thresholds is somewhat more involved, but can be treated similarly.  However, as the number of trials is generated by a memoryless process, observers gain no further information about how many trials remain after the present, and will thus not adjust their thresholds as in Section~\ref{S:dynamic.multi}. 

Our conclusions also depend on the assumptions  the observer makes about the environment.
For instance, an observer may not know the exact probability of change between trials, so that $\ep^{\rm assumed} \neq \ep^{\rm true}$.  Indeed, there is evidence that human subjects adjust their estimate of $\ep$ across a sequence of trials, and assume trials are correlated, even when they are not~\citep{cho02,gao09,Yu08}.  Our model can be expanded to a hierarchical model in which both the states $H^{1:n}$ and the change rate, $\ep$, are inferred across a sequence of trials~\citep{radillo17}.  This would result in model ideal observers that use the current estimate of the $\ep$ to set the initial bias on a trial~\citep{Yu08}.  A similar approach could also allow us to model observers that incorrectly learn, or make wrong assumptions, about the rate of change. 

Modeling approaches can also point to the neural computations that underlie the decision-making process. Cortical representations of previous trials have been identified~\citep{nogueira17,akrami18}, but it is unclear how the brain learns and combines latent properties of the environment (the change rate $\ep$), with sensory information to make inferences. Recent work on repetition bias in working memory suggests short-term plasticity serves to encode priors in neural circuits, which bias neural activity patterns during memory retention periods~\citep{papadimitriou15,kilpatrick18}. Such synaptic mechanisms may also allow for the encoding of previously accumulated information in sequential decision-making tasks, even in cases where state sequences are not correlated. Indeed, evolution may imprint characteristics of natural environments onto neural circuits, shaping  whether and how $\ep$ is learned in experiments and influencing trends in human response statistics~\citep{fawcett14,glaze18}. Ecologically adapted decision rules  may be difficult to train away, especially if they do not impact performance too adversely~\citep{todd07}.

The normative model we have derived can be used to identify task parameter ranges that will help tease apart the assumptions made by subjects~\citep{cho02,gao09,goldfarb12}. Since our model describes the behavior of an ideal observer, it can be used to determine whether experimental subjects use near-optimal or sub-optimal evidence-accumulation strategies. Furthermore, our adaptive DDM  opens a window to further instantiations of the DDM that consider the influence of more complex state-histories on the evidence-accumulation strategies adopted within a trial. Uncovering common assumptions about state-histories will help guide future experiments, and help us better quantify the biases and core mechanisms of human decision-making.

\section*{Acknowledgements}
This work was supported by an NSF/NIH CRCNS grant (R01MH115557). KN was supported by an NSF Graduate Research Fellowship. ZPK was also supported by an NSF grant (DMS-1615737). KJ was also supported by NSF grant DBI-1707400 and DMS-1517629.

\appendix

\section{Derivation of the drift-diffusion model for a single trial}
\label{DDM_app}
We assume that an optimal observer integrates a stream of noisy measurements $\xi_{1:s}^1 = (\xi_1^1, \xi_2^1, ..., \xi_s^1)$ at equally spaced times $t_{1:s} = (t_1, t_2, ..., t_s)$~\citep{wald48,bogacz06}. The likelihood functions, $f_{\pm}(\xi) : = \P (\xi |H_{\pm}),$ define the probability of each measurement, $\xi$, conditioned on the environmental state, $H_{\pm}$. Observations in each trial are combined with the prior probabilities $P(H_{\pm})$. We assume a symmetric prior, $P(H_{\pm}) = 1/2$. The probability ratio is then
\begin{align}
R_s^1 = \frac{\P(H^1 = H_+| \xi_{1:s}^1)}{\P(H^1 = H_-| \xi_{1:s}^1)} = \frac{f_+(\xi_1^1) f_+(\xi_2^1) \cdots f_+(\xi_s^1) \P (H^1 = H_+)}{f_-(\xi_1^1) f_-(\xi_2^1) \cdots f_-(\xi_s^1) \P(H^1 = H_-)}, \label{LR1}
\end{align}
due to the independence of each measurement $\xi_s^1$. Thus, if  $R_s^1> 1 (R_s^1<1)$ then $H^1 = H_{+} (H^1 = H_{-})$ is the more likely state. Eq.~(\ref{LR1}) can be written recursively
\begin{align}
R_s^1 = \left( \frac{f_+(\xi_s^1)}{f_-(\xi_s^1)} \right) R_{s-1}^1,  \label{probrat}
\end{align}
where $R_0^1 = \P(H_+)/\P(H_-) = 1$, due to the assumed prior at the beginning of trial 1.

Taking the logarithm of Eq.~(\ref{probrat}) allows us to express the recursive relation for the log-likelihood ratio (LLR) $L_s^1 = \ln R_s^1$ as an iterative sum
\begin{align}
L_s^1 = L_{s-1}^1 + \ln \frac{f_+(\xi_s^1)}{f_-(\xi_s^1)},
\end{align}
where $L_0^1 = \ln \left[ \P(H_+)/\P(H_-) \right]$, so if $L_s^1 \gtrless 0$ then $H^1 = H_{\pm}$ is the more likely state. Taking the continuum limit $\Delta t \to 0$ of the timestep $\Delta t : = t_s - t_{s-1}$, we can use the functional central limit theorem to yield the DDM (See \citet{bogacz06,velizcuba16} for details):
\begin{align}
\d y^1 &= g^1 \d t + \sqrt{2 D} \d W, 
\end{align}
where $W$ is a Wiener process, the drift $g^1 \in g_{\pm} = \frac{1}{\Delta t} {\rm E}_{\xi} \left[ \ln \frac{f_+(\xi)}{f_-(\xi)} | H_{\pm} \right]$ depends on the state, and $2 D = \frac{1}{\Delta t} {\rm Var}_{\xi} \left[  \ln \frac{f_+(\xi)}{f_-(\xi)} | H_{\pm} \right]$ is the variance which depends only on the noisiness of each observation, but not the state.

With Eq.~(\ref{ddm1}) in hand, we can relate $y^1(t)$ to the probability of either state $H_{\pm}$ by noting its Gaussian statistics in the case of free boundaries 
\begin{align*}
\P(y^1(t) | H^1 = H_{\pm}) = \frac{1}{\sqrt{4 \pi D t}} \e^{-(y^1(t) \mp t)^2/(4D t)},
\end{align*}
so that
\begin{align}
 \frac{\P(H^1 = H_+ | y^1(t))}{\P(H^1 = H_- | y^1(t))} = \frac{\P(y^1(t) | H^1 = H_+)}{\P(y^1(t)| H^1 = H_-)} = \e^{y^1(t)/D}.  \label{ytoLR}
\end{align}
Note, an identical relation is obtained in \citet{bogacz06} in the case of absorbing boundaries. Either way, it is clear that $y^1 = D \cdot {\rm LLR}_1$. This means that before any observations have been made $\e^{y^1(0)/D} = \frac{\P(H^1 = H_+)}{\P(H^1 = H_-)}$, so we scale the log ratio of the prior by $D$ to yield the initial condition $y_0^1$ in Eq.~(\ref{ddm1}).

\section{Threshold determines probability of being correct}
\label{c2thresh}
We note that Eq.~(\ref{ytoLR}) extends to any trial $j$ in the case of an ideal observer. In this case, when the decision variable $y^2(T_2) = \theta$, we can write
\begin{align}
\frac{\P(H^2 = H_+ | y^2(T_2) = \theta )}{\P(H^2 = H_- | y^2(T_2) = \theta)}  = \e^{\theta/D}, \label{y2thet}
\end{align}
so that a rearrangement of Eq.~(\ref{y2thet}) yields
\begin{align}
c_2 : = \frac{\P (H^2 = H_+ | y^2(T_2) = \theta) \P (y^2(T_2) = \theta) }{ \P ( d_2 = + 1)} = \frac{1}{1 + \e^{- \theta/ D}} = c_1, \label{c2c1}
\end{align}
Therefore the probability of correct decisions on trials 1 and 2 are equal, and determined by $\theta = \theta_{1:2},$ and $D.$ A similar computation shows that the threshold, $\theta_j = \theta$, and diffusion rate, $D$, \emph{but not the initial belief}, determine the probability of a correct response on any trial $j$.

To demonstrate self-consistency in the general case of decision thresholds that may differ between trials $\theta_{1:2}$, we can also derive the formula for $c_2$ in the case $y_0^2 \leq \theta_2$. Our alternative derivation of Eq.~\eqref{c2c1} begins by computing the total probability of a correct choice on trial 2 by combining the conditional probabilities of a correct choice given decision $d^1 = +1$ leading to initial belief $y_0^2$, and 
decision $d^1 = -1$ leading to initial belief $-y_0^2$:
\begin{align}
c_2 =& \left[ (1- \ep) c_1 + \ep (1-c_1) \right] \cdot \pi_{\theta_2}(y_0^2) + \left[ \ep c_1 + (1-\ep) (1-c_1) \right] \cdot \pi_{\theta_2}(-y_0^2), \nonumber \\
=& \frac{(1 - \ep) \e^{\theta_1/D} + \ep}{\e^{\theta_1/D} + 1} \frac{(1- \ep)( \e^{\theta_1/D} - \e^{- \theta_2/D}) + \ep (1 - \e^{(\theta_1- \theta_2)/D})}{((1 - \ep) \e^{\theta_1/D} + \ep)( 1- \e^{- 2\theta_2/D})} \nonumber \\
& + \frac{\ep \e^{\theta_1/D} + (1-\ep)}{\e^{\theta_1/D} + 1} \frac{\ep ( \e^{\theta_1/D} - \e^{- \theta_2/D}) + (1- \ep)(1 - \e^{(\theta_1- \theta_2)/D}) }{( \ep \e^{\theta_1/D} + (1- \ep))(1 - \e^{- 2 \theta_2/D})} \nonumber \\
=&  \frac{( \ep \e^{\theta_1 /D} + (1 - \ep)) ((1- \ep) \e^{\theta_1/D} + \ep) (\e^{\theta_1/D} + 1) (1 - \e^{- \theta_2/D})}{( \ep \e^{\theta_1 /D} + (1 - \ep)) ((1- \ep) \e^{\theta_1/D} + \ep) (\e^{\theta_1/D} + 1) (1 - \e^{- 2 \theta_2/D})} \nonumber \\
c_2 =& \frac{1}{1 + \e^{- \theta_2/D}}, 
\end{align}
showing again that the correct probability $c_2$ on trial 2 is solely determined by $\theta_2$ and $D$, assuming $y_0^2 \leq \theta_2$.

\section{Mean response time on the second trial} \label{avgtime}
We can compute the average time until a decision on trial 2 by marginalizing over  $d_1 = \pm 1$, 
\begin{align}
DT_2 | (H^2 = H_{\pm}) =& \sum_{s = \pm 1} T (s \cdot y_0^2; \theta_2) \cdot \P (d_1 = s | H^2 = H_{\pm}), \label{DT2Hp}
\end{align}
since $d_1 = \pm 1$ will lead to the initial condition $\pm y_0^2$ in trial 2. We can compute $T(\pm y_0^2; \theta)$ using Eq.~(\ref{Tygen}) as
\begin{align}
T(\pm y_0^2; \theta_2) =& \theta_2 \left[ \coth \frac{\theta_2}{D} - {\rm csch} \frac{\theta_2}{D} \left[ \frac{\ep \e^{\theta_1/D} + (1- \ep)}{(1- \ep) \e^{\theta_1/D} + \ep} \right]^{\pm 1} \right] \mp D \ln \frac{(1- \ep) \e^{\theta_1/D} + \ep}{\ep \e^{\theta_1/D} + (1-\ep)}. \label{genT2}
\end{align}
The second terms in the summands of Eq.~(\ref{DT2Hp}) are given by
\begin{align*}
\P (d_1 = \pm 1 | H^2 = H_{\pm}) &= \frac{\P (H^2 = H_{\pm} | d_1 = \pm 1) \P (d_1 = \pm 1) }{\P (H^2  = H_\pm)} = (1- \ep) c_1 + \ep (1-c_1), \\
\P (d_1 = \mp 1 | H^2 = H_{\pm}) &= \frac{\P (H^2 =  H_{\pm} | d_1 = \mp 1) \P (d_1 = \mp 1) }{\P (H^2  = H_{\pm})} = \ep c_1  + (1-\ep) (1-c_1),
\end{align*}
and note $DT_2 = \left[ DT_2 | H^2 = H_+ \right] \P(H^2 = H_+) + \left[ DT_2 | H^2 = H_- \right] \P(H^2 = H_-) = DT_2 | (H^2 = H_{\pm})$.

Using Eqs.~(\ref{DT2Hp}-\ref{genT2}) and simplifying assuming $\theta_{1:2} = \theta$, we obtain Eq.~\eqref{DT2}. For unequal thresholds, $\theta_1 \neq \theta_2$, a similar computation yields Eq.~\eqref{E:avgdyn2}.

\section{Repetition-dependence of correct probabilities and decision times}
\label{rep_app}
Here we derive formulas for correct probabilities $c_j$ and average decision times $DT_j$ in trial $j$, conditioned on whether the current trial is a repetition ($R$: $H^j = H^{j-1}$) or alternation ($A$: $H^j \ne H^{j-1}$)  compared with the previous trial. To begin, we first determine the correct probability, conditioned on repetitions ($R$), $c_j|R := c_j | (H^j = H^{j-1} = H_{\pm})$:
\begin{align*}
c_j | R =&  \sum_{s = \pm 1} \pi_{\pm \theta}(\pm s \cdot y_0^j) \cdot \P (d_{j-1} = \pm s | H^j = H^{j-1} = H_{\pm})   \\
=& \frac{1 - \e^{- \theta/D} \e^{- y_0^j/D}}{1 - \e^{-2 \theta/D}}  \cdot  \frac{1}{1 + \e^{- \theta/D}} + \frac{1 - \e^{- \theta/D} \e^{+y_0^j/D}}{1 - \e^{-2 \theta/D}}  \cdot \frac{\e^{- \theta/D}}{1 + \e^{- \theta/D}} \\
=& \frac{(1 - \ep) \e^{\theta/D}}{ \left[ 1 + \e^{- \theta/D} \right] \left[ (1 - \ep) \e^{\theta/D} + \ep \right]} + \frac{\ep}{\left[ 1 + \e^{- \theta/D} \right] \left[ \ep \e^{\theta/D} + (1 - \ep) \right]} \\
= & \frac{(1 - \ep)\ep \e^{2 \theta/D}+ (1- \ep) \e^{\theta/D} + \ep^2}{ \left[ 1 + \e^{- \theta/D} \right] \left[ (1 - \ep) \e^{\theta/D} + \ep \right] \left[ \ep \e^{\theta/D} + (1 - \ep) \right]} .
\end{align*}
In the case of alternations, $c_j | A := c_j | (H^j \neq H^{j-1}) = c_j | (H^j  \neq H^{j-1} = H_\pm) $:
\begin{align*}
c_j | A=&  \sum_{s = \pm 1} \pi_{\mp \theta}(\mp s \cdot y_0^j) \cdot \P (d_{j-1} = \mp s | H^j \neq H^{j-1} = H_{\pm})   \\
=& \frac{1 - \e^{- \theta/D} \e^{- y_0^j/D}}{1 - \e^{-2 \theta/D}}  \cdot  \frac{\e^{- \theta/D}}{1 + \e^{- \theta/D}} + \frac{1 - \e^{- \theta/D} \e^{+y_0^j/D}}{1 - \e^{-2 \theta/D}}  \cdot \frac{1}{1 + \e^{- \theta/D}} \\
=& \frac{(1 - \ep)}{ \left[ 1 + \e^{- \theta/D} \right] \left[ (1 - \ep) \e^{\theta/D} + \ep \right]} + \frac{\ep \e^{\theta/D}}{\left[ 1 + \e^{- \theta/D} \right] \left[ \ep \e^{\theta/D} + (1 - \ep) \right]} \\
= & \frac{(1 - \ep)\ep \e^{2 \theta/D}+ \ep \e^{\theta/D} + (1-\ep)^2}{ \left[ 1 + \e^{- \theta/D} \right] \left[ (1 - \ep) \e^{\theta/D} + \ep \right] \left[ \ep \e^{\theta/D} + (1 - \ep) \right]}.
\end{align*}
Thus, we can show that the correct probability in the case of repetitions is higher than that for alternations, $c_j | R > c_j | A$ for any $\theta > 0$ and $\ep \in [0, 0.5)$, since
\begin{align*}
c_j | R -  c_j | A= \frac{(1- 2\ep) \left[ \e^{\theta/D} - 1\right]}{ \left[ 1 + \e^{- \theta/D} \right] \left[ (1 - \ep) \e^{\theta/D} + \ep \right] \left[ \ep \e^{\theta/D} + (1 - \ep) \right]} > 0.
\end{align*}

Now, we also demonstrate that decision times are shorter for repetitions than for alternations. First, note that when the current trial is a repetition, we can again marginalize to compute $DT_j |R : = DT_j| (H^j = H^{j-1} = H_{\pm})$:
\begin{align*}
DT_j | R &= T(+y_0^j; \theta) \frac{1}{1 + \e^{- \theta/D}} + T(-y_0^j; \theta) \frac{\e^{- \theta/D}}{1 + \e^{- \theta/D}},
\end{align*}
whereas in the case of alternations
\begin{align*}
DT_j | A &= T(+y_0^j; \theta) \frac{\e^{- \theta/D}}{1 + \e^{- \theta/D}} + T(-y_0^j; \theta) \frac{1}{1 + \e^{- \theta/D}}.
\end{align*}
By subtracting $DT_j|A$ from $DT_j|R$, we can show that $DT_j|R < DT_j|A$ for all $\theta > 0$ and $\ep \in [0, 0.5)$. To begin, note that
\begin{align*}
DT_j | R - DT_j | A = \frac{1 - \e^{- \theta/D}}{1 + \e^{- \theta/D}} \left[ T(+y_0^j; \theta) - T(-y_0^j; \theta) \right].
\end{align*}
Since it is clear the first term in the product above is positive for $\theta, D>0$, we can show $DT_j | R - DT_j | A<0$ and thus $DT_j | R < DT_j | A $ if $T(+y_0^j; \theta) < T(-y_0^j; \theta)$. We verify this is so by first noting the function $F(y) = (\e^{y/D}-\e^{-y/D})/y$ is always nondecreasing since $F'(y) = \left[ \frac{y}{D} \cosh \frac{y}{D} - \sinh \frac{y}{D} \right]/y^2 \geq 0$ since $y/D \geq \tanh y/D$. In fact $F(y)$ is strictly increasing everywhere but when $y=0$. This means that as long as $y_0 < \theta$, $F(y_0)< F(\theta)$, which implies
\begin{align*}
T(+y_0^j ; \theta) - T(-y_0^j; \theta) = 2 \theta \frac{\e^{y_0/D} - \e^{-y_0/D}}{\e^{\theta/D} - \e^{- \theta/D}} - 2 y_0 &< 0,
\end{align*}
so $T(+y_0^j ; \theta) < T(-y_0^j; \theta)$ as we wished to show, so $DT_j | R < DT_j | A $.

Moving to three state sequences, we note that we can use our formulas for $c_j|R$ and $c_j|A$ in~\ref{rep_app} as well as $\pi_{\theta}(y)$ described by Eq.~(\ref{genpi}) to compute
\begin{align*}
c_{RR}: = c_j|RR &= \pi_{\pm \theta} (\pm y_0^j) \cdot c_j|R + \pi_{\pm \theta} (\mp y_0^j) \cdot (1-c_j)|R, \\
c_{RA}: = c_j|RA & = \pi_{\pm \theta} (\mp y_0^j) \cdot c_j|R + \pi_{\pm \theta} (\pm y_0^j) \cdot (1-c_j)|R, \\
c_{AR}: = c_j|AR &= \pi_{\pm \theta} ( \pm y_0^j) \cdot c_j|A + \pi_{\pm \theta} (\mp y_0^j) \cdot (1-c_j)|A, \\
c_{AA}: = c_j|AA &= \pi_{\pm \theta} ( \mp y_0^j) \cdot c_j|A + \pi_{\pm \theta} (\pm y_0^j) \cdot (1-c_j)|A.
\end{align*}
Applying similar logic to the calculation of the decision times as they depend on the three state sequences, we find that
\begin{align*}
T_{RR} : = DT_j| RR &= T(+y_0^j; \theta) \cdot c_j|R + T(-y_0^j; \theta) \cdot (1-c_j)|R, \\
T_{RA}: = DT_j |RA & = T(- y_0^j; \theta) \cdot c_j|R + T(+y_0^j; \theta) \cdot (1-c_j)|R, \\
T_{AR}: = DT_j |AR &= T(+ y_0^j; \theta)  \cdot c_j|A + T(-y_0^j; \theta) \cdot (1-c_j)|A, \\
T_{AA} : = DT_j |AA &= T(-y_0^j; \theta) \cdot c_j|A + T(+y_0^j; \theta) \cdot (1-c_j)|A.
\end{align*}

\section{Numerical simulations}
All numerical simulations of the DDM were performed using the Euler-Maruyama method with a timestep $dt = 0.005$, computing each point from $10^5$ realizations. To optimize reward rates, we used the Nelder-Mead simplex method ({\tt fminsearch} in MATLAB). All code used to generate figures will be available on {\tt github}.

\bibliographystyle{elsarticle-harv}

\end{document}